\documentclass[prd,superscriptaddress,amsfonts,amssymb,amsmath,showpacs,twocolumn]{revtex4-2}
\usepackage{bm}
\usepackage{amsfonts}
\usepackage{latexsym}
\usepackage[latin1]{inputenc}
\usepackage{graphicx}
\usepackage{amsmath}
\usepackage{palatino}
\usepackage{mathpazo}
\usepackage{textcomp}
\usepackage{float}
\usepackage{booktabs}
\usepackage{dcolumn}
\usepackage{hyperref}
\hypersetup{colorlinks,citecolor=blue}
\usepackage{amsmath}
\usepackage{xcolor}
\usepackage{orcidlink}
\usepackage[caption=false]{subfig}
\usepackage{commath}
\def\jnl@style{\it}
\def\aaref@jnl#1{{\jnl@style#1}}

\def\aaref@jnl#1{{\jnl@style#1}}

\def\aj{\aaref@jnl{AJ}}                   
\def\apj{\aaref@jnl{ApJ}}                 
\def\apjl{\aaref@jnl{ApJ}}                
\def\apjs{\aaref@jnl{ApJS}}               
\def\apss{\aaref@jnl{Ap\&SS}}             
\def\aap{\aaref@jnl{A\&A}}                
\def\aapr{\aaref@jnl{A\&A~Rev.}}          
\def\aaps{\aaref@jnl{A\&AS}}              
\def\mnras{\aaref@jnl{Mon.~Not.~Roy.~Astron.~Soc.}}             
\def\prd{\aaref@jnl{Phys.~Rev.~D}}        
\def\prc{\aaref@jnl{Phys.~Rev.~C}}  
\def\prl{\aaref@jnl{Phys.~Rev.~Lett.}}    
\def\qjras{\aaref@jnl{QJRAS}}             
\def\skytel{\aaref@jnl{S\&T}}             
\def\ssr{\aaref@jnl{Space~Sci.~Rev.}}     
\def\zap{\aaref@jnl{ZAp}}                 
\def\nat{\aaref@jnl{Nature}}              
\def\aplett{\aaref@jnl{Astrophys.~Lett.}} 
\def\apspr{\aaref@jnl{Astrophys.~Space~Phys.~Res.}} 
\def\physrep{\aaref@jnl{Phys.~Rep.}}      
\def\physscr{\aaref@jnl{Phys.~Scr}}       
\def\commat{\aaref@jnl{Comm.~Math.~Phys.}}              
\def\science{\aaref@jnl{Science}}               
\def\cqg{\aaref@jnl{Classical Quant.~Grav.}}            
\def\jpcs{\aaref@jnl{JPCS}}                                     
\def\ijmpd{\aaref@jnl{Int.~J.~Mod.~Phys.~D}}                    
\def\grg{\aaref@jnl{Gen.~Relat.~Gravit.}}               
\def\rpp{\aaref@jnl{Rep.~Prog.~Phys.}}          
\def\npa{\aaref@jnl{Nucl.~Phys.~A}}        
\def\lrr{\aaref@jnl{Living Rev.~Rel.}}                   
\def\jcap{\aaref@jnl{J.~Cosmology Astropart.~Phys.}}    
\def\rmp{\aaref@jnl{Rev.~Mod.~Phys.}}   
\def\epjc{\aaref@jnl{Eur.~Phys.~J.~C}}

\allowdisplaybreaks[1]

\begin{document}

\color{black}       

\title{Accelerating Universe scenario in anisotropic $f\left( Q\right) $ cosmology}
\author{M. Koussour\orcidlink{0000-0002-4188-0572}}
\email{pr.mouhssine@gmail.com}
\affiliation{Quantum Physics and Magnetism Team, LPMC, Faculty of Science Ben
M'sik,\\
Casablanca Hassan II University,
Morocco.}

\author{M. Bennai\orcidlink{0000-0003-1424-7699}}
\email{mdbennai@yahoo.fr }
\affiliation{Quantum Physics and Magnetism Team, LPMC, Faculty of Science Ben
M'sik,\\
Casablanca Hassan II University,
Morocco.} 
\affiliation{Lab of High Energy Physics, Modeling and Simulations, Faculty of
Science,\\
University Mohammed V-Agdal, Rabat, Morocco.}
%
\date{\today}
\begin{abstract}
In this paper, we investigate the exact solution of an anisotropic
space-time in the context of $f(Q)$ gravity, where $f\left( Q\right) $ is the
arbitrary function of the non-metricity scalar $Q$. Here, we consider a
specific power-law form as $f\left( Q\right) =\alpha Q^{n+1}+\beta $, where $\alpha $, $\beta $, and $n$ are free model parameters. Using power-law
cosmology ($a\left( t\right) \propto t^{m}$), we analyze the physical
behavior of cosmological parameters such as the energy density, pressure,
EoS parameter, skewness parameter. Further, we validated the model with energy conditions and found that our $f\left( Q\right) $\
cosmological model behaves like the quintessence model in the present and $%
\Lambda $CDM in the future.
\end{abstract}

\maketitle

\date{\today}
\section{Introducion}

One of the most common problems in modern cosmology is the problem of
the current expansion of the Universe. General Relativity (GR) as the theoretical framework of modern cosmology says that the expansion of the Universe is decelerating, but astrophysical observational data have a different opinion: the current Universe is in an acceleration phase \cite{Obs1,
Obs2, Obs3, Obs4, Obs5, Obs6, Obs7, Obs8, Obs9, Obs10, Obs11, Obs12}. The
phenomenon of cosmic inflation and the problem of initial singularity is
other subjects of discussion \cite{BB1, BB2}. Before Hubble showed through
observations that our universe is expanding, cosmologists believed that our
universe is stable and static, so Einstein added a small positive
cosmological constant ($\Lambda $) to his field equations to go with
this idea, after which Einstein described that the cosmological constant is
the biggest mistake in his life. In the late 1990s, it was shown that the
cosmological constant was returned because it might be a suitable candidate
for Dark Energy (DE). DE is a strange form of energy that has negative
pressure $(p)$ and positive energy density $(\rho)$ and is assumed to be
responsible for the cosmic acceleration. DE has a negative equation of state
(EoS) parameter $(\omega =\frac{p}{\rho }<0)$, and we can classify other DE
models by this parameter, i.e. quintessence $-1<\omega <-0.33$, cosmological
constant $\omega =-1$, phantom DE $\omega <-1$ \cite{Ratra, Sami1, Sami2,
Peebles}.

Another proposal to explain the current acceleration of the Universe comes
from modified theories of gravity (MTG) \cite{myrz1} . One of the most
recently developed examples is the $f\left( Q\right) $ theory of gravity or
the symmetric teleparallel gravity (STG) \cite{Jimenez1}. This theory is a
generalized version of teleparallel gravity \cite{myrz2} in which gravity is caused to
the non-metricity and where $f\left( Q\right) $ is an arbitrary function of
the non-metricity scalar $Q$. In this theory, the geometry is torsion and
curvature free and the non-metricity term $Q$ only determines the
gravitational interactions. The STG is demonstrated in the so-called
coincidence gauge by assuming that the connection is symmetric \cite{Xu}.
The geometric basis of GR is Riemannian geometry, while gravity $f\left(
Q\right) $ whose basis is Weyl geometry which is a generalization of
Riemannian geometry \cite{myrz3, myrz5}. Also Weyl geometry is a particular case of Weyl-Cartan geometry in which torsion disappears \cite{myrz4}. The interesting about this
theory is that it could explain the current acceleration of the Universe
like other MTG without using DE \cite{Koussour1}.
Many works have been done in this context so far. The first cosmological
solutions in $f\left( Q\right) $ gravity are discussed in \cite{Jimenez2}.
Mandal et al. presented a complete analysis of energy conditions for $f\left(
Q\right) $ gravity models and constraint families of $f\left( Q\right) $
models compatible with the present accelerating expansion of the Universe 
\cite{Mandal1}. In addition, the cosmography in $f\left( Q\right) $ gravity
is considered by Mandal et al. \cite{Mandal2}. The growth index of matter
perturbations has been examined in the context of $f\left( Q\right) $
gravity in \cite{Harko1} and, several other issues in $f\left( Q\right) $
gravity are discussed in \cite{Dimakis, Shekh, Hassan}.

According to the cosmological principle, isotropic models of the Universe are most suitable for studying the structure of the Universe on a large
scale. Nevertheless, cosmologists believe that the early universe may not
have been completely uniform. Further, the theoretical arguments and the
anomalies observed in the CMB (Cosmic Microwave Background) are clear
evidence of the existence of an anisotropic phase, which was later called
the isotropic phase \cite{Planck2015, Planck2018}. This forecasting
motivates us to create cosmological models in the early phases of the
Universe with models that have an anisotropic background. Hence, the
presence of anisotropy in the early phases of the Universe is an intriguing
topic to study. There are models in cosmology that respect these criteria
called Bianchi models, a Bianchi type-I is one such anisotropic cosmological
model that is considered a generalization of the flat Friedmann-Lema\^{\i}%
tre-Robertson-Walker (FLRW) model. Therefore, in this work, we will explore
Bianchi type-I models in the context of $f(Q)$ gravity because it is an
interesting and still very new topic. Exact solutions of the symmetric
teleparallel gravity field equations for a locally rotationally symmetric
(LRS) Bianchi type-I space-time were studied in \cite{De}. Solutions of a
Bianchi type-I space-time were examined in \cite{Koussour2} using the hybrid
expansion law.

This manuscript is organized as follows: In Sec \ref{sec2}, we presented a
brief description of the mathematical formalism of $f(Q)$ gravity. In Sec. %
\ref{sec3}, the field equations in anisotropic space-time are evaluated for $%
f\left( Q\right) $ gravity. In Sec. \ref{sec4}, we have discussed the energy
conditions in $f(Q)$ gravity. In Sec. \ref{sec5}, we consider the specific
power-law form of $f(Q)$ gravity as $f\left( Q\right) =\alpha Q^{n+1}+\beta $, where 
$\alpha $, $\beta $, and $n$ are free model parameters. Further, we discussed
the behavior of several cosmological parameters in the same section. Lastly,
we discussed our conclusions in Sec. \ref{sec6}.

\section{Brief description of the mathematical formalism of $f(Q)$ gravity}

\label{sec2}

In Weyl-Cartan geometry, the so-called symmetric metric tensor $g_{\mu \nu }$
can be considered as a generalization of the gravitational potential, and it
is fundamentally used to define the length of a vector, and we also need an
asymmetric connection $\Sigma {^{\gamma }}_{\mu \nu }$ to defines the
covariant derivatives and parallel transport. Thus, the general affine
connection can be decomposed into three elements: the Christoffel symbol ${%
\Gamma ^{\gamma }}_{\mu \nu }$, the contortion tensor ${C^{\gamma }}_{\mu
\nu }$, and the disformation tensor ${L^{\gamma }}_{\mu \nu }$,
respectively, which is given by \cite{Xu}

\begin{equation}
\Sigma {^{\gamma }}_{\mu \nu }={\Gamma ^{\gamma }}_{\mu \nu }+{C^{\gamma }}%
_{\mu \nu }+{L^{\gamma }}_{\mu \nu },  \label{eqn1}
\end{equation}%
where the Levi-Civita connection ${\Gamma ^{\gamma }}_{\mu \nu }$ of the
metric $g_{\mu \nu }$ has the form

\begin{equation}
{\Gamma ^{\gamma }}_{\mu \nu }\equiv \frac{1}{2}g^{\gamma \sigma }\left( 
\frac{\partial g_{\sigma \nu }}{\partial x^{\mu }}+\frac{\partial g_{\sigma
\mu }}{\partial x^{\nu }}-\frac{\partial g_{\mu \nu }}{\partial x^{\sigma }}%
\right) ,  \label{eqn2}
\end{equation}%
the contorsion tensor ${C^{\gamma }}_{\mu \nu }$ can be written as

\begin{equation}
{C^{\gamma }}_{\mu \nu }\equiv \frac{1}{2}{T^{\gamma }}_{\mu \nu }+T_{(\mu
}{}^{\gamma }{}_{\nu )},  \label{eqn3}
\end{equation}%
where ${T^{\gamma }}_{\mu \nu }\equiv 2{\Sigma ^{\gamma }}_{[\mu \nu ]}$ in
Eq. (\ref{eqn3}) is the torsion tensor. Finally, the disformation tensor ${%
L^{\gamma }}_{\mu \nu }$ is derived from the non-metricity tensor $Q_{\gamma
\mu \nu }$ as

\begin{equation}
{L^{\gamma }}_{\mu \nu }\equiv \frac{1}{2}g^{\gamma \sigma }\left( Q_{\nu
\mu \sigma }+Q_{\mu \nu \sigma }-Q_{\gamma \mu \nu }\right) .  \label{eqn4}
\end{equation}

In the above equation, the non-metricity tensor $Q_{\gamma \mu \nu }$ is
specific as the (minus) covariant derivative of the metric tensor with
regard to the Weyl-Cartan connection $\Sigma {^{\gamma }}_{\mu \nu }$, i.e. $%
Q_{\gamma \mu \nu }=\nabla _{\gamma }g_{\mu \nu }$, and it can be obtained

\begin{equation}
Q_{\gamma \mu \nu }=-\frac{\partial g_{\mu \nu }}{\partial x^{\gamma }}%
+g_{\nu \sigma }\Sigma {^{\sigma }}_{\mu \gamma }+g_{\sigma \mu }\Sigma {%
^{\sigma }}_{\nu \gamma }.  \label{eqn5}
\end{equation}

The connection is presumed to be torsionless and curvatureless within the
current background. It corresponds to the pure coordinate transformation
from the trivial connection mentioned in \cite{Jimenez1}. Thus, for a flat
and torsion-free connection, the connection (\ref{eqn1}) can be
parameterized as

\begin{equation}
\Sigma {^{\gamma }}_{\mu \beta }=\frac{\partial x^{\gamma }}{\partial \xi
^{\rho }}\partial _{\mu }\partial _{\beta }\xi ^{\rho }.  \label{eqn6}
\end{equation}

Now, $\xi ^{\gamma }=$ $\xi ^{\gamma }\left( x^{\mu }\right) $ is an
invertible relation. It is always possible to get a coordinate system so
that the connection $\Sigma {^{\gamma }}_{\mu \nu }$ vanish. This condition
is called coincident gauge and has been used in many studies of symmetric
teleparallel equivalent to GR (STEGR) and in this condition the covariant\
derivative $\nabla _{\gamma }$ reduces to the partial derivative $\partial
_{\gamma }$ \cite{Xu}. Thus, in the coincident gauge coordinate, we get

\begin{equation}
Q_{\gamma \mu \nu }=-\partial _{\gamma }g_{\mu \nu }.  \label{eqn7}
\end{equation}

The STG is a geometric description of gravity equivalent to GR within
coincident gauge coordinates in which $\Sigma _{\mu \nu }=0$ and ${C^{\gamma
}}_{\mu \nu }=0$, and consequently from Eq. (\ref{eqn1}) we can conclude
that \cite{Xu}

\begin{equation}
{\Gamma ^{\gamma }}_{\mu \nu }=-{L^{\gamma }}_{\mu \nu }.  \label{eqn8}
\end{equation}

The $f\left( Q\right) $ theory of gravity is a generalization or
modification of the symmetric teleparallel equivalent of GR. The action for
this theory is given \cite{Jimenez1, Jimenez2, Mandal1, Mandal2}

\begin{equation}
S=\int \left[ \frac{1}{2\kappa }f(Q)+\mathcal{L}_{m}\right] d^{4}x\sqrt{-g},
\label{eqn9}
\end{equation}%
where $\kappa =8\pi G=1$, $f(Q)$ can be expressed as the arbitrary function
of non-metricity scalar $Q$, $g$ is the determinant of the metric tensor $%
g_{\mu \nu }$,\ and $\mathcal{L}_{m}$ is the matter Lagrangian density. Now,
the non-metricity tensor $Q_{\gamma \mu \nu }$ and its traces can be written
as

\begin{equation}
Q_{\gamma \mu \nu }=\nabla _{\gamma }g_{\mu \nu }\,,  \label{eqn10}
\end{equation}%
\begin{equation}
Q_{\gamma }={{Q_{\gamma }}^{\mu }}_{\mu }\,,\qquad \widetilde{Q}_{\gamma }={%
Q^{\mu }}_{\gamma \mu }\,.  \label{eqn11}
\end{equation}

In addition, the superpotential tensor (non-metricity conjugate) can be
expressed as 
\begin{equation}
4{P^{\gamma }}_{\mu \nu }=-{Q^{\gamma }}_{\mu \nu }+2Q_{({\mu ^{^{\gamma }}}{%
\nu })}-Q^{\gamma }g_{\mu \nu }-\widetilde{Q}^{\gamma }g_{\mu \nu }-\delta _{%
{(\gamma ^{^{Q}}}\nu )}^{\gamma }\,,  \label{eqn12}
\end{equation}%
where the trace of the non-metricity tensor can be obtained as 
\begin{equation}
Q=-Q_{\gamma \mu \nu }P^{\gamma \mu \nu }\,.  \label{eqn13}
\end{equation}

Now, the matter energy-momentum tensor is defined as

\begin{equation}
T_{\mu \nu }=-\frac{2}{\sqrt{-g}}\frac{\delta (\sqrt{-g}\mathcal{L}_{m})}{%
\delta g^{\mu \nu }}\,.  \label{eqn14}
\end{equation}

By varying the modified Einstein-Hilbert action (\ref{eqn9}) with respect to
the metric tensor $g_{\mu \nu }$, the gravitational field equations obtained
as 
\begin{widetext}
\begin{equation}
\frac{2}{\sqrt{-g}}\nabla _{\gamma }\left( \sqrt{-g}f_{Q}P^{\gamma }{}_{\mu
\nu }\right) -\frac{1}{2}fg_{\mu \nu }+f_{Q}\left( P_{\mu \gamma i}Q_{\nu
}{}^{\gamma i}-2Q_{\gamma i\mu }P^{\gamma i}{}_{\nu }\right) =T_{\mu \nu },
\label{eqn15}
\end{equation}%
\end{widetext}where $f_{Q}=\frac{df}{dQ}$ and $\nabla _{\gamma }$\ denotes
the covariant derivative.

\section{$f\left( Q\right) $ field equations in anisotropic space-time}

\label{sec3}

The standard FLRW Universe is isotropic and homogeneous. Hence, to address
the anisotropic nature of the Universe in $f\left( Q\right) $ gravity, which
manifests as anomalies found in the CMB, the Bianchi type-I Universe is
indeed important because it represents a spatially homogeneous, but not
isotropic. Thus, we consider a Bianchi-type I space-time in the form 
\begin{equation}
ds^{2}=-dt^{2}+A^{2}(t)dx^{2}+B^{2}(t)(dy^{2}+dz^{2}),  \label{eqn16}
\end{equation}%
where metric potentials $A\left( t\right) $ and $B\left( t\right) $ depend
only on cosmic time $\left( t\right) $. Here, to complete the choice of the
anisotropic type space-time, the equation of state (EoS) parameter of the
gravitational fluid must also be generalized, and from another point of
view, to give a more reasonable model, an anisotropic nature must be
exhibited as described in \cite{Sahoo}. Thus, the energy-momentum tensor for
the anisotropic fluid can be expressed as 
\begin{align}
T_{\nu }^{\mu }& =\text{diag}(-\rho ,p_{x},p_{y},p_{z})\,,  \label{eqn17} \\
& =\text{diag}(-1,\omega _{x},\omega _{y},\omega _{z})\rho ,  \notag \\
& =\text{diag}(-1,\omega ,(\omega +\delta ),(\omega +\delta ))\rho ,  \notag
\end{align}%
where $\rho $ is the energy density of the anisotropic fluid, $p_{x}$, $%
p_{y} $, $p_{z}$ are the pressures and $\omega _{x}$, $\omega _{y}$, $\omega
_{z}$\ are the directional EoS parameters along $x$, $y$ and $z$ coordinates
respectively. The deviation from isotropy is parametrized by setting $\omega
_{x}=\omega $ and then introducing the deviations along $y$ and $z$ axes by
the skewness parameter $\delta $, where $\omega $ and $\delta $ are
functions of cosmic time only \cite{De}.

For anisotropic fluid as matter contents, the corresponding field equations
of Bianchi type-I space-time are obtained as \cite{De} 
\begin{widetext}
\begin{equation}
\frac{f}{2}+f_{Q}\left[ 4\frac{\dot{A}}{A}\frac{\dot{B}}{B}+2\left( \frac{%
\dot{B}}{B}\right) ^{2}\right]  =\rho,  \label{eqn18}
\end{equation}

\begin{equation}
\frac{f}{2}-f_{Q}\left[ -2\frac{\dot{A}}{A}\frac{\dot{B}}{B}-2\frac{\ddot{B}%
}{B}-2\left( \frac{\dot{B}}{B}\right) ^{2}\right] +2\frac{\dot{B}}{B}\dot{Q}%
f_{QQ} =-\omega \rho, \label{eqn19}
\end{equation}

\begin{equation}
\frac{f}{2}-f_{Q}\left[ -3\frac{\dot{A}}{A}\frac{\dot{B}}{B}-\frac{\ddot{A}}{%
A}-\frac{\ddot{B}}{B}-\left( \frac{\dot{B}}{B}\right) ^{2}\right] +\left( 
\frac{\dot{A}}{A}+\frac{\dot{B}}{B}\right) \dot{Q}f_{QQ} =-(\omega +\delta
)\rho,  \label{eqn20}
\end{equation}%
\end{widetext}where the dot $\left( \text{\textperiodcentered }\right) $
denotes the ordinary differentiation with regard to cosmic time $\left(
t\right) $. The corresponding non-metricity scalar can be written as

\begin{equation}
Q=-2\left( \frac{\overset{.}{B}}{B}\right) ^{2}-4\frac{\overset{.}{A}}{A}%
\frac{\overset{.}{B}}{B}.  \label{eqn21}
\end{equation}

In order to simplify the form of the field equations (\ref{eqn18})-(\ref%
{eqn20}) and write them in terms of the non-metricity scalar $Q$, the
directional Hubble parameters $H_{x}=\frac{\overset{.}{A}}{A},H_{y}=\frac{%
\overset{.}{B}}{B}$ and average Hubble parameter $H=\frac{1}{3}\left(
H_{x}+2H_{y}\right) $, we use the following relations: $\frac{\partial }{%
\partial t}\left( \frac{\dot{A}}{A}\right) =\frac{\ddot{A}}{A}-\left( \frac{%
\dot{A}}{A}\right) ^{2}$ and $Q=-2H_{y}^{2}-4H_{x}H_{y}$. The field
equations (\ref{eqn18})-(\ref{eqn20}) becomes 
\begin{equation}
\frac{f}{2}-Qf_{Q}=\rho ,  \label{eqn22}
\end{equation}%
\begin{equation}
\frac{f}{2}+2\frac{\partial }{\partial t}\left[ H_{y}f_{Q}\right]
+6Hf_{Q}H_{y}=-\omega \rho ,  \label{eqn23}
\end{equation}%
\begin{equation}
\frac{f}{2}+\frac{\partial }{\partial t}\left[ f_{Q}(H_{x}+H_{y})\right]
+3Hf_{Q}\left( H_{x}+H_{y}\right) =-(\omega +\delta )\rho .  \label{eqn24}
\end{equation}

Using Eqs. (\ref{eqn22}) and (\ref{eqn23}), the EoS parameter ($\omega =%
\frac{p}{\rho }$) is 
\begin{equation}
\omega =\frac{-1}{\frac{f}{2}-Qf_{Q}}\left[ \frac{f}{2}+2\frac{\partial }{%
\partial t}\left[ H_{y}f_{Q}\right] +6Hf_{Q}H_{y}\right] .  \label{eqn25}
\end{equation}

Similarly, using Eqs. (\ref{eqn22}), (\ref{eqn23}) and (\ref{eqn24}) we find
the skewness parameters as 
\begin{equation}
\delta =\frac{1}{\frac{f}{2}-Qf_{Q}}\left[ \frac{\partial }{\partial t}\left[
f_{Q}\left( H_{y}-H_{x}\right) \right] +3Hf_{Q}\left( H_{y}-H_{x}\right) %
\right] .  \label{eqn26}
\end{equation}

The above parameters are useful enough to check the validity of the
constructed cosmological model, and the reason for expressing them in terms
of average Hubble parameter and directional Hubble parameters is that
understanding the behavior of the Universe will be more convenient. This
means that the volumetric expansion law will remain in a proper form. In 
\cite{De} the study of anisotropic cosmological models in $f(Q)$ gravity has
shown that the solutions can follow the power law using the anisotropy
relation ($\theta ^{2}\propto \sigma ^{2}$). Motivated by this work, we will
study here the volumetric power law expansion of the form $V=AB^{2}=t^{m}$,
where $m$ be the arbitrary constant that is chosen according to the
observational constraints. In the following sections, we will analyze our
cosmological models with these considerations.

\section{Energy conditions}

\label{sec4}

The energy conditions (ECs) are a set of conditions that describe the
geodesics of the Universe. In this background, the ECs are intended to
verify the existence of a phase of accelerating expansion of the Universe.
Like these condition can be obtained from the well-known Raychaudhury
equations, whose forms are \cite{Raychaudhuri, Nojiri, Ehlers, Capozziello1}

\begin{equation}
\frac{d\theta }{d\tau }=-\frac{1}{3}\theta ^{2}-\sigma _{\mu \nu }\sigma
^{\mu \nu }+\omega _{\mu \nu }\omega ^{\mu \nu }-R_{\mu \nu }u^{\mu }u^{\nu
},  \label{eqn27}
\end{equation}

\begin{equation}
\frac{d\theta }{d\tau }=-\frac{1}{2}\theta ^{2}-\sigma _{\mu \nu }\sigma
^{\mu \nu }+\omega _{\mu \nu }\omega ^{\mu \nu }-R_{\mu \nu }n^{\mu }n^{\nu
}.  \label{eqn28}
\end{equation}

Here, $\theta $ is the expansion factor, $n^{\mu }$ is the null vector, and $%
\sigma ^{\mu \nu }$ and $\omega _{\mu \nu }$ are, respectively, the shear
and the rotation connected with the vector field $u^{\mu }$. For
mathematical details of deriving the Raychaudhury equations in Weyl geometry
with the existence of non-metricity, see these references \cite{Mandal1,
Arora}. For attractive gravity, Eqs. (\ref{eqn27}) and (\ref{eqn28}) fulfill
the following conditions

\begin{equation}
R_{\mu \nu }u^{\mu }u^{\nu }\geq 0,  \label{eqn29}
\end{equation}

\begin{equation}
R_{\mu \nu }n^{\mu }n^{\nu }\geq 0.  \label{eqn30}
\end{equation}

Hence, if we consider the Universe has a perfect fluid matter distribution,
the ECs for $f\left( Q\right) $ gravity are given by

\begin{itemize}
\item Weak energy conditions (WEC) if $\rho \geq 0,\rho +p\geq 0$;

\item Null energy condition (NEC) if $\rho +\rho \geq 0$;

\item Dominant energy conditions (DEC) if $\rho \geq 0,\left\vert
p\right\vert \leq \rho $.

\item Strong energy conditions (SEC) if $\rho +3p\geq 0$.
\end{itemize}

\section{Cosmological $f\left( Q\right) $\ model}

\label{sec5}

To study the anisotropic nature of the Universe, we consider the specific
power law form of $f(Q)$ i.e.

\begin{equation}
f\left( Q\right) =\alpha Q^{n+1}+\beta ,  \label{eqn31}
\end{equation}%
where $\alpha $, $\beta $, and $n$ are free model parameters. This specific
functional form for $f(Q)$ was motivated by the work in \cite{Koussour1}.
For $n=0$ we find the simplest linear form of $f(Q)$ function i.e. $f\left(
Q\right) =\alpha Q+\beta $ and for $n=1$ the model takes the non-linear form
i.e. $f\left( Q\right) =\alpha Q^{2}+\beta $. In addition, to reduce the
number of unknowns, we presumed an anisotropic relation between the
directional scale factors in the form $A=B^{k}$ where $k\left( \neq
0,1\right) $ is an arbitrary real number. For a power-law background, the
directional scale factors can be derived as: $A\left( t\right) =t^{\frac{km}{%
k+2}}$, $B\left( t\right) =t^{\frac{m}{k+2}}$. As a result, the directional
Hubble parameters are given as: $H_{x}=\left( \frac{km}{k+2}\right) \frac{1}{%
t}$, $H_{y}=\left( \frac{m}{k+2}\right) \frac{1}{t}$. The above anisotropy
relation is chosen on the basis of astronomical observations associated with
the velocity redshift relation for extragalactic sources which indicate that
the Hubble expansion of the Universe may attain isotropy when $\frac{\sigma 
}{\theta }$ is constant \cite{Kantowski, Collins, Shamir, Koussour3,
Koussour4}.

Now, using Eq. (\ref{eqn31}) we find the following expressions for energy
density, pressure, EoS parameter, and skewness parameter, respectively.

\begin{widetext}
\begin{equation}
\rho =\frac{1}{2}\left( \beta +\alpha 2^{n+1}\psi ^{n+1}-\alpha
2^{n+2}(n+1)\psi ^{n+1}\right)   \label{eqn32}
\end{equation}

\begin{equation}
p=-\frac{-\alpha (k+2)m2^{n+2}\left( 2n^{2}+3n+1\right) \psi ^{n}+\alpha
m^{2}2^{n+1}(2(k+2)n+3)\psi ^{n}+\beta (k+2)^{2}t^{2}}{2(k+2)^{2}t^{2}}
\label{eqn33}
\end{equation}

\begin{equation}
\omega =\frac{\alpha (k+2)m2^{n+2}\left( 2n^{2}+3n+1\right) \psi ^{n}+\alpha
m^{2}\left( -2^{n+1}\right) (2(k+2)n+3)\psi ^{n}-\beta (k+2)^{2}t^{2}}{%
\alpha (2k+1)m^{2}2^{n+1}(2n+1)\psi ^{n}+\beta (k+2)^{2}t^{2}}  \label{eqn34}
\end{equation}

\begin{equation}
\delta =\frac{\alpha (k-1)(k+2)m2^{n+1}(n+1)(-m+2n+1)\psi ^{n}}{\alpha
(2k+1)m^{2}2^{n+1}(2n+1)\psi ^{n}+\beta (k+2)^{2}t^{2}}  \label{eqn35}
\end{equation}%
\end{widetext}where $\psi \left( t\right) =-\frac{(2k+1)m^{2}}{(k+2)^{2}t^{2}%
}$.

\begin{figure}[tbp]
\centerline{\includegraphics[scale=0.7]{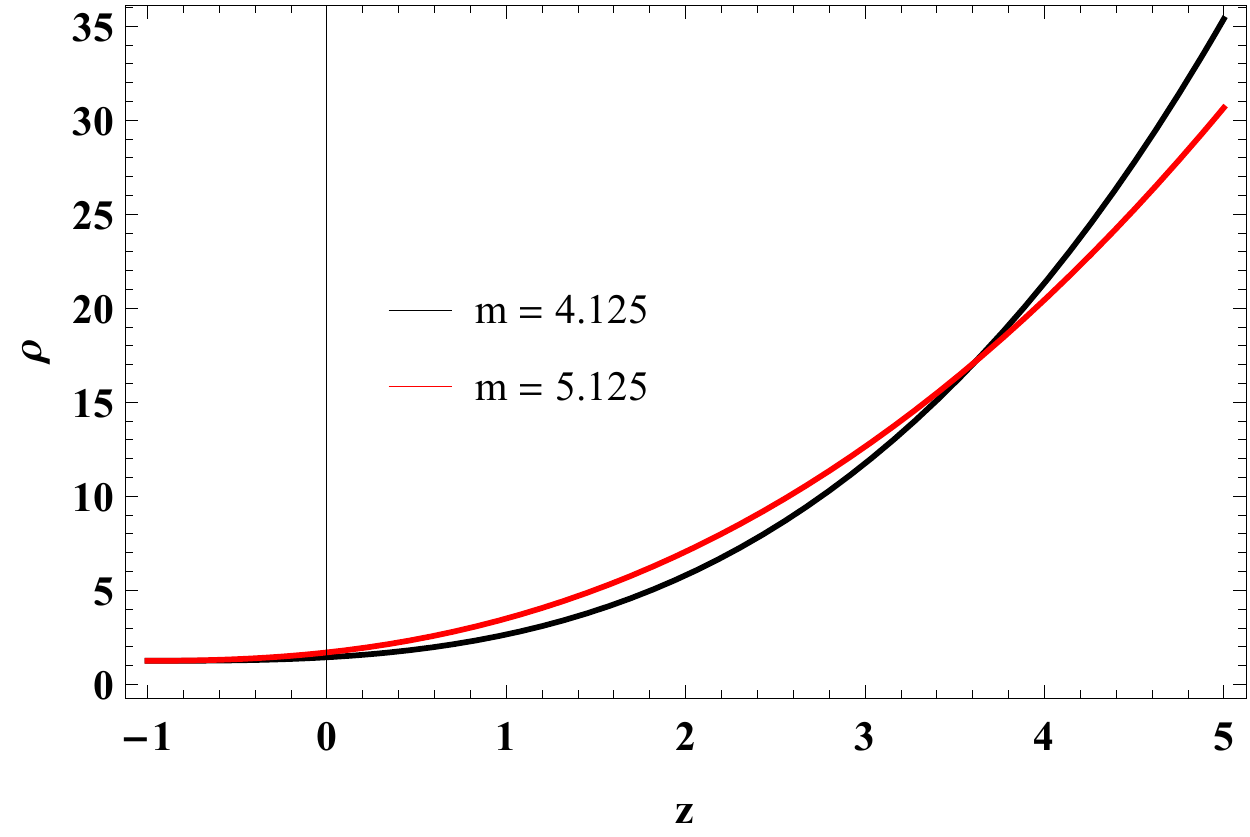}}
\caption{Plot of cosmic energy density ($\protect\rho $) versus redshift ($z$%
)\ for $\protect\alpha =-0.001$, $\protect\beta =2.5$, $n=1$ and $k=0.64$.}
\label{fig1}
\end{figure}

\begin{figure}[h]
\centerline{\includegraphics[scale=0.7]{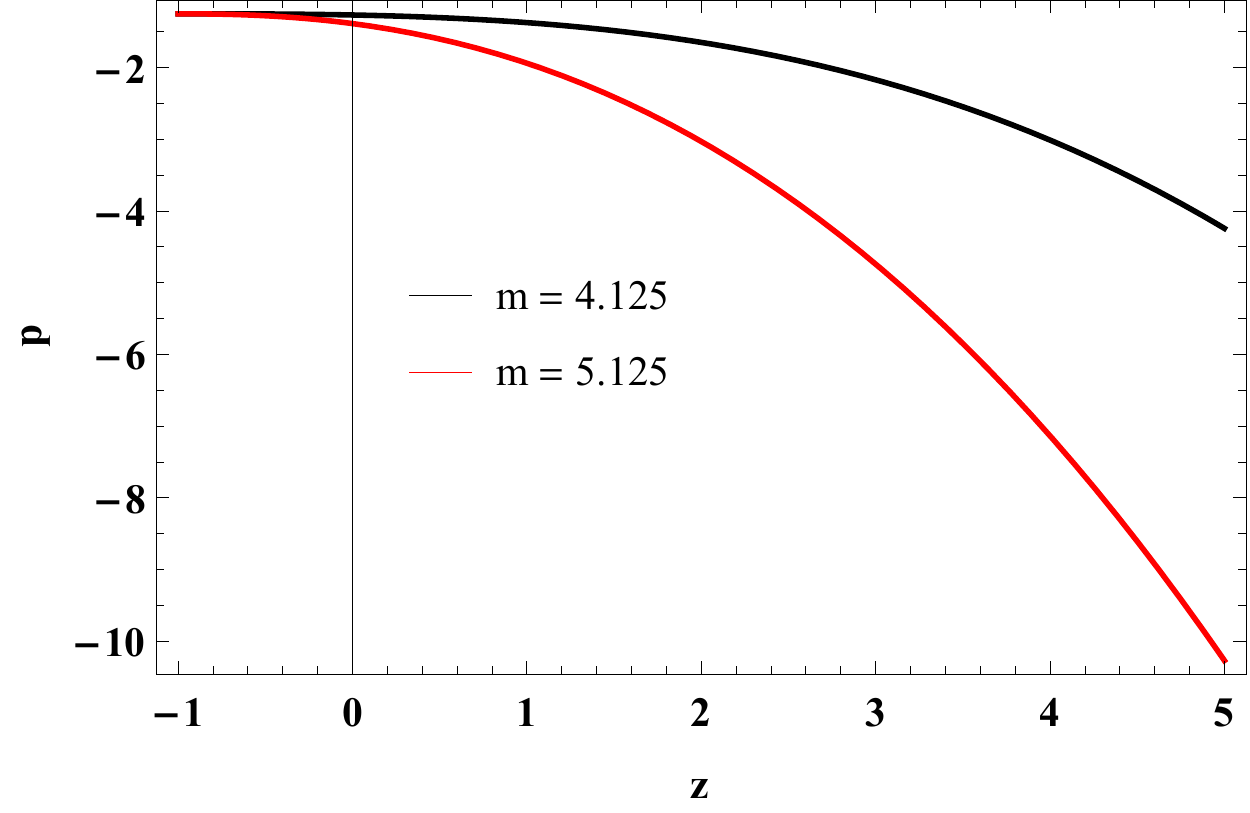}}
\caption{Plot of cosmic pressure ($p$) versus redshift ($z$)\ for $\protect%
\alpha =-0.001$, $\protect\beta =2.5$, $n=1$ and $k=0.64$.}
\label{fig2}
\end{figure}

\begin{figure}[h]
\centerline{\includegraphics[scale=0.7]{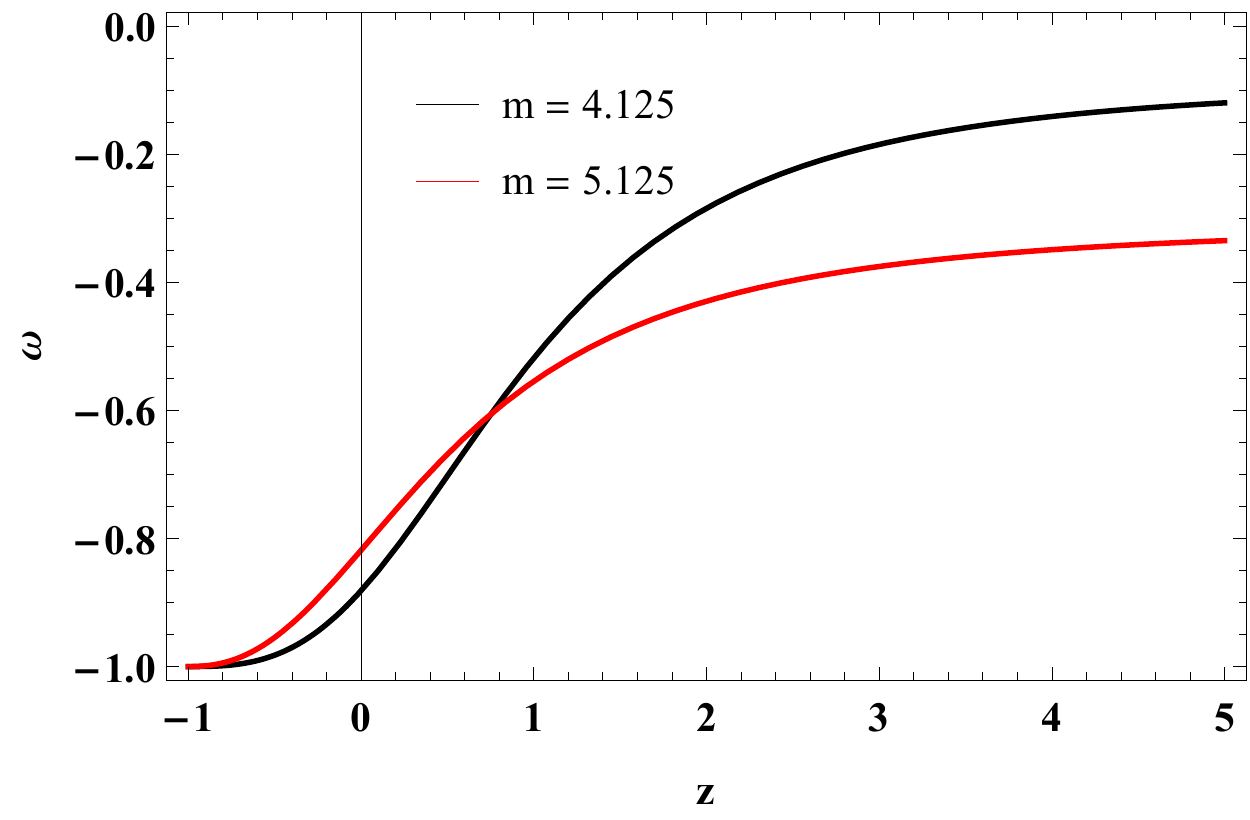}}
\caption{Plot of EoS parameter ($\protect\omega $) versus redshift ($z$)\
for $\protect\alpha =-0.001$, $\protect\beta =2.5$, $n=1$ and $k=0.64$.}
\label{fig3}
\end{figure}

\begin{figure}[h]
\centerline{\includegraphics[scale=0.7]{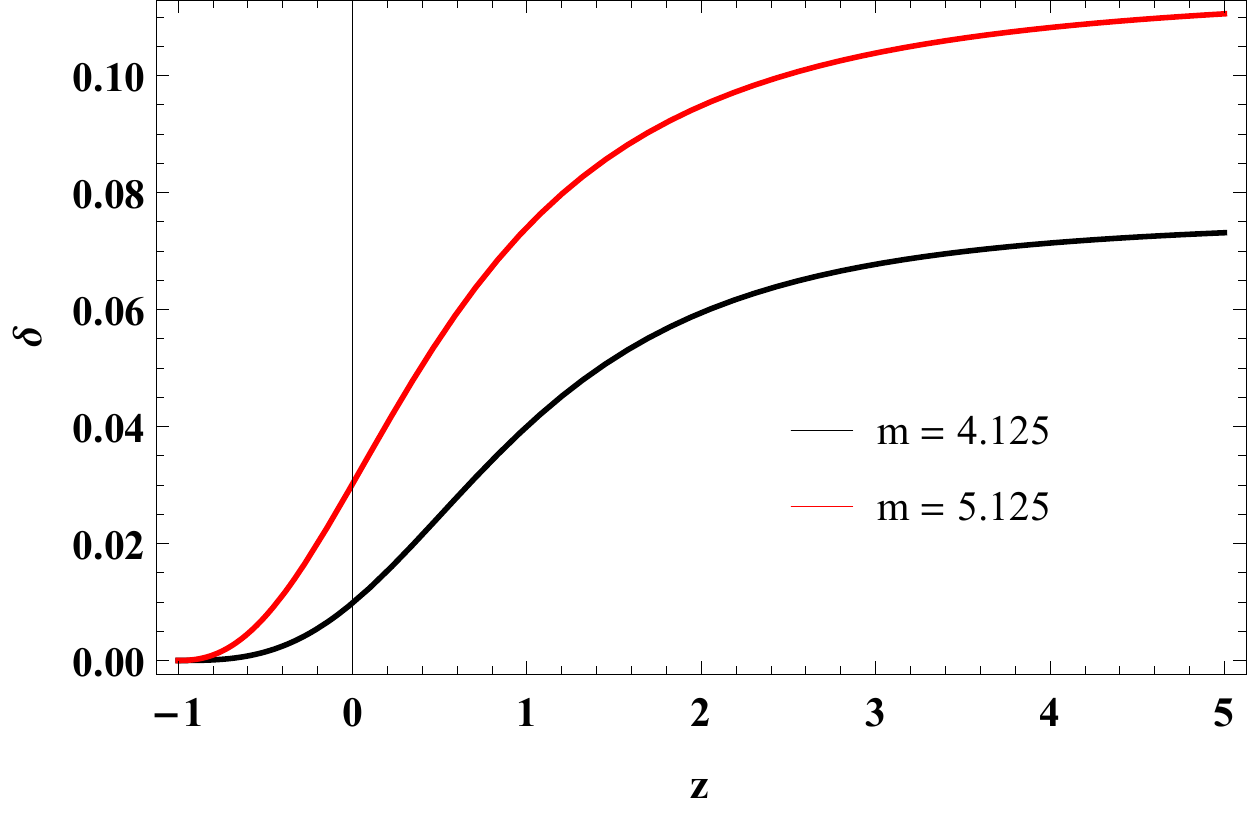}}
\caption{Plot of skewness parameter ($\protect\delta $) versus redshift ($z$%
)\ for $\protect\alpha =-0.001$, $\protect\beta =2.5$, $n=1$ and $k=0.64$.}
\label{fig4}
\end{figure}

First, all cosmological parameters ($\rho ,$ $p,$ $\omega ,$ $\delta $) are
plotted in terms of cosmic redshift ($z$) with the help of the time-redshift
relation $t\left( z\right) =\left( 1+z\right) ^{-\frac{3}{m}}$ for the
different values of $m=4.125$ and $5.125$. From Fig. \ref{fig1} it is clear
that the cosmic energy density remains positive throughout the evolution of
the Universe and is an increasing function of cosmic redshift. It starts
with a positive value and approaches zero at $z\rightarrow -1$ while Fig. %
\ref{fig2} show that the cosmic pressure is a decreasing function of cosmic
redshift, which starts with large negative values and tends to zero at $z=0$
(present epoch) and $z\rightarrow -1$ (future). Such behavior of cosmic
pressure is caused by the current cosmic acceleration, or so-called DE in
the context of GR. Also, from Fig. \ref{fig3} it can be observed that the
EoS parameter is similar to the behavior of quintessence DE at
present epoch and converges to the cosmological constant ($\Lambda $) at $%
z=-1$. Further, Fig. \ref{fig4} indicates that the skewness parameter starts
with a positive value in the past and tends to zero at $z\rightarrow -1$.
Hence, the Universe in our model transforms from the phase of anisotropy to
the phase of isotropy.

Using Eqs. (\ref{eqn32}) and (\ref{eqn33}) in the ECs, we obtain

\begin{widetext}
\begin{equation}
NEC\Leftrightarrow \frac{%
\alpha m2^{n+1}(n+1)\left( k(m+2n+1)-m+4n+2\right)
\psi ^{n}}{(k+2)^{2}t^{2}%
}\geq 0  \label{eqn36}
\end{equation}

\begin{equation}
DEC\Leftrightarrow \frac{-\alpha (k+2)m2^{n+1}\left( 2n^{2}+3n+1\right)
\psi
^{n}+\alpha m^{2}2^{n+1}(3kn+k+3n+2)\psi ^{n}+\beta (k+2)^{2}t^{2}}{%
(k+2)^{2}t^{2}}\geq 0  \label{eqn37}
\end{equation}

\begin{equation}
SEC\Leftrightarrow -\frac{-3\alpha (k+2)m2^{n+1}\left(
2n^{2}+3n+1\right)
\psi ^{n}+\alpha m^{2}2^{n+1}\left( k(n-1)+5n+4\right)
\psi ^{n}+\beta
(k+2)^{2}t^{2}}{(k+2)^{2}t^{2}}\geq 0  \label{eqn38}
\end{equation}
\end{widetext}

\begin{figure}[h]
\centerline{\includegraphics[scale=0.7]{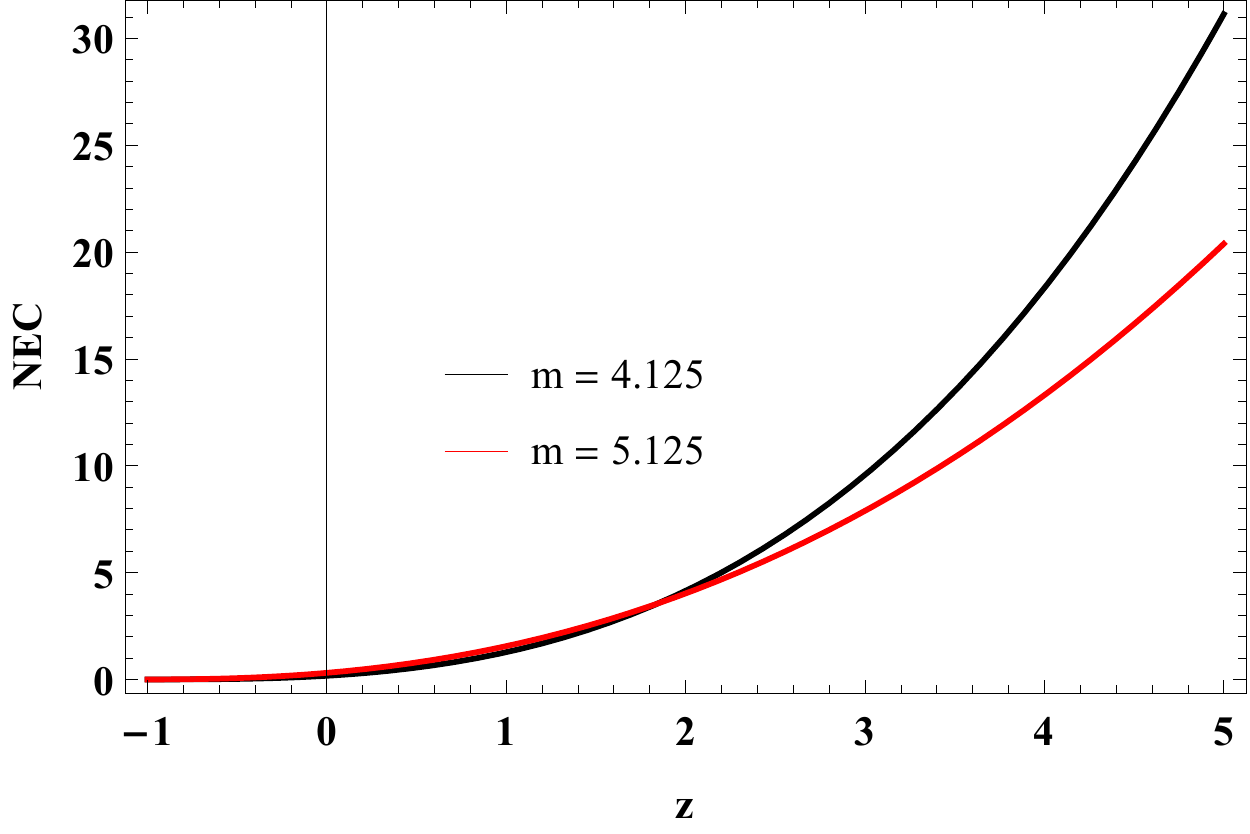}}
\caption{Plot of the NEC versus redshift ($z$)\ for $\protect\alpha =-0.001$%
, $\protect\beta =2.5$, $n=1$ and $k=0.64$.}
\label{fig5}
\end{figure}

\begin{figure}[h]
\centerline{\includegraphics[scale=0.7]{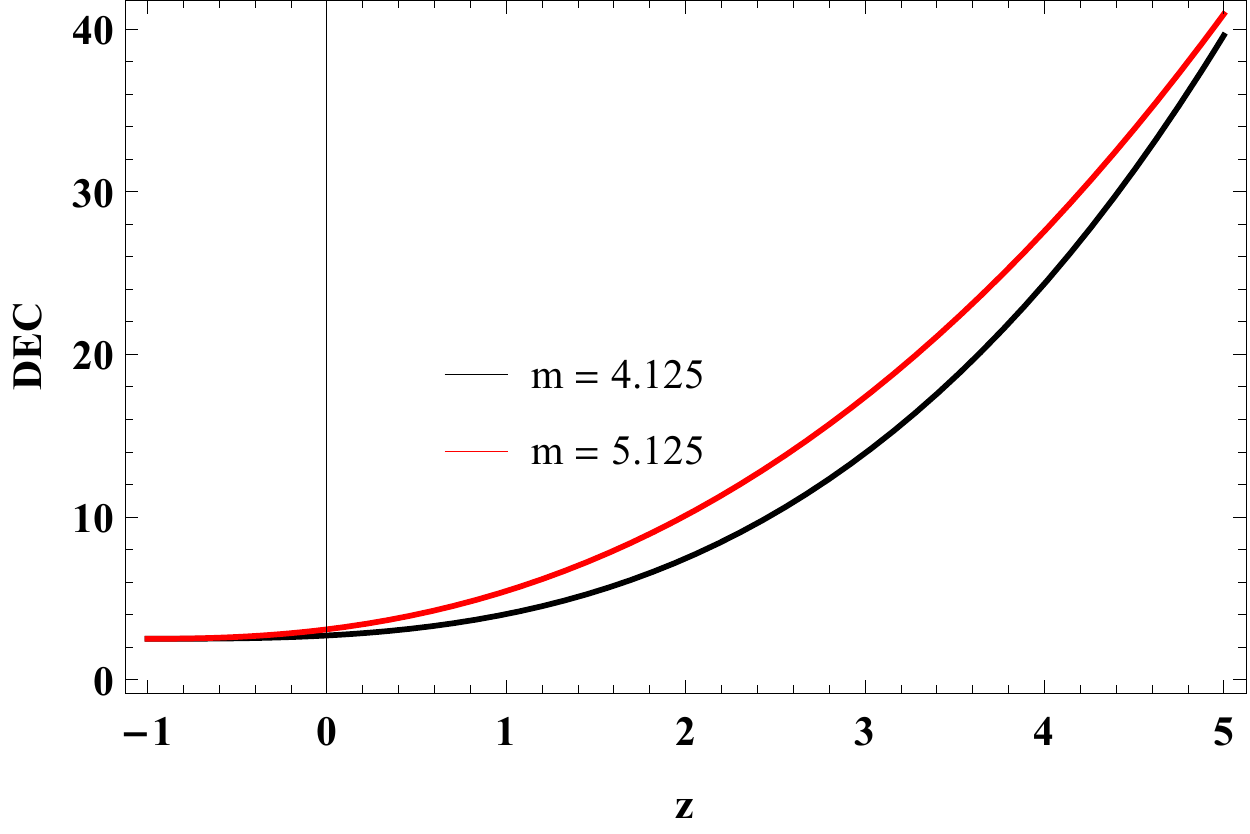}}
\caption{Plot of the DEC versus redshift ($z$)\ for $\protect\alpha =-0.001$%
, $\protect\beta =2.5$, $n=1$ and $k=0.64$.}
\label{fig6}
\end{figure}

\begin{figure}[h]
\centerline{\includegraphics[scale=0.7]{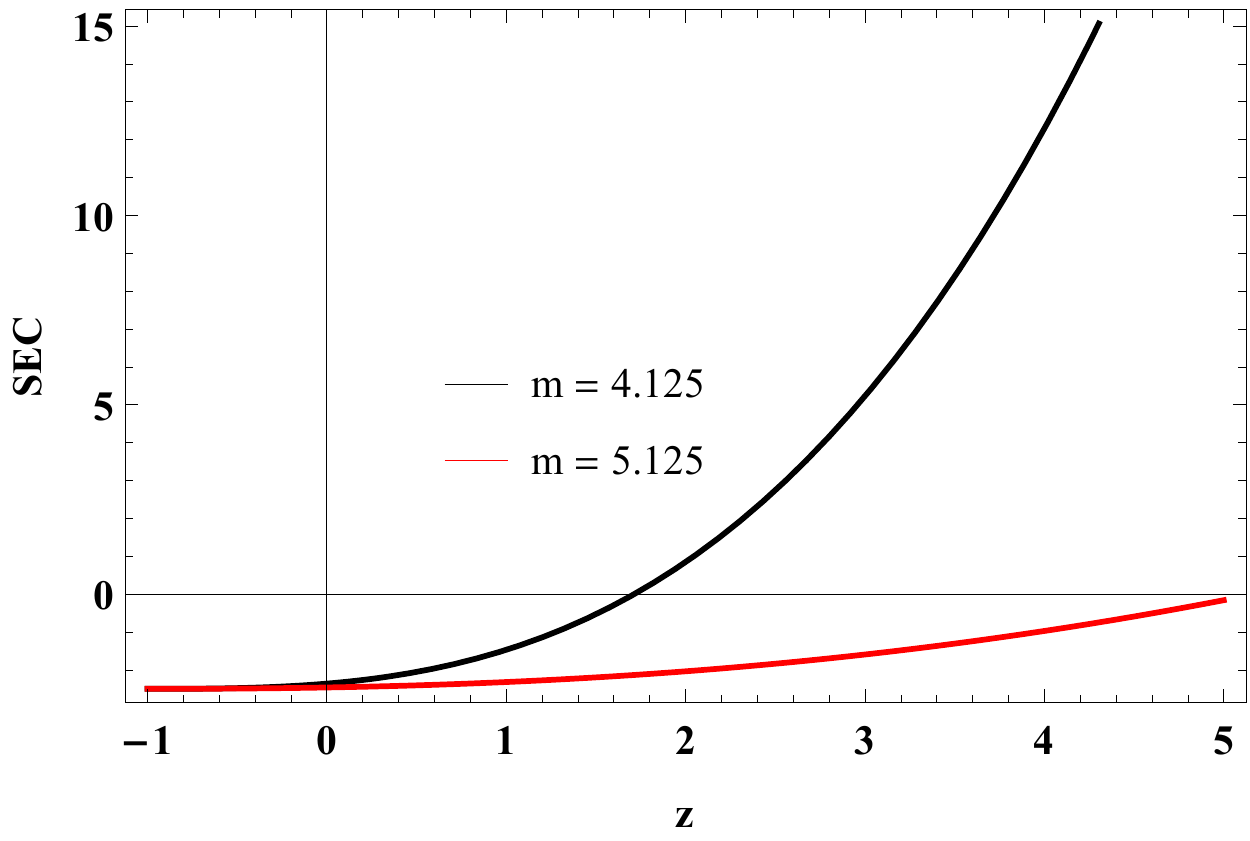}}
\caption{Plot of the SEC versus redshift ($z$)\ for $\protect\alpha =-0.001$%
, $\protect\beta =2.5$, $n=1$ and $k=0.64$.}
\label{fig7}
\end{figure}

From Figs. \ref{fig5}-\ref{fig7} it is observed that all the ECs (WEC, NEC
and DEC) fulfill while the SEC is violated for this model. This behavior is
consistent with the current scenario of the Universe.

\section{Discussions and conclusions}

\label{sec6}

In the context of the modified theories of gravity (MTG), the current cosmic
acceleration can be explained by introducing new terms into the field
equations, which means that we do not need to add another form of energy or
matter to cause this acceleration. Generally, the current acceleration phase
can be predicted by several cosmological parameters which can be directly
measured by astronomical observations such as the equation of state (EoS)
parameter and deceleration parameter. According to recent observational data
of astronomy and cosmology such as WMAP satellites which combined data from
the $H_{0}$ measurements, supernovae, CMB (Cosmic Microwave Background), and
BAO (Baryonic Acoustic Oscillations), the current value of the EoS parameter
is $\omega _{0}=-1.084\pm 0.063$. Further, in the year 2015 Planck
collaboration showed that $\omega _{0}=-1.006\pm 0.0451$ \cite{Planck2015}
and furthermore, in 2018 it informed that $\omega _{0}=-1.028\pm 0.032$ \cite%
{Planck2018}. 

In this paper, we investigated the current acceleration of the universe in
the context of the recently proposed $f(Q)$ modified gravity. For this
purpose, we considered a specific power law form as $f\left( Q\right)
=\alpha Q^{n+1}+\beta $, where $\alpha $, $\beta $, and $n$ are free model
parameters and examine the exact solutions of the Bianchi type-I
cosmological model. To reduce the number of unknowns, we presumed an
anisotropic relation between the directional scale factors in the form $%
A=B^{k}$, where $k\left( \neq 0,1\right) $ is an arbitrary real number. In
the context of the power-law cosmology ($a\left( t\right) \propto t^{m}$),
we found the expressions for energy density, pressure, EoS parameter and
skewness parameter for our cosmological model. Further, we checked the
energy conditions of our model. The cosmic energy energy remains positive
throughout the evolution of the Universe and is an increasing function of
cosmic redshift. It starts with a positive value and approaches zero in the
future, while the cosmic pressure is a decreasing function of cosmic
redshift, which starts with large negative values and tends to zero in the
present and future. In addition, we observed that, for all $m$ values the
EoS parameter is similar to the behavior of quintessence DE at present epoch
and converges to the cosmological constant in the future. The current values
of EoS parameter corresponding to $m=4.125$ and $5.125$ are $\omega
_{0}=-0.8809$ and $\omega _{0}=-0.8159$, respectively. As a result, it can
be said that these values are in agreement with the above values obtained
from the observational data. Further, the skewness parameter starts with a
positive value in the past and tends to zero in the future. Thus, the
Universe in our model transforms from the phase of anisotropy to the phase
of isotropy. All the energy conditions fulfill while the SEC is violated for
this model. The violation of SEC leads directly to the current acceleration
phase of the universe.

Other cosmological parameters that are no less important than the ones above
are the Hubble parameter and deceleration parameter. In cosmology, the
Hubble parameter represents the expansion rate of the universe, while the
deceleration parameter is a measure of the variation in the expansion of the
Universe, if $q<0$ the Universe is in a phase of accelerated expansion and
if $q>0$ the Universe is in a phase of decelerated expansion. In the context
of power-law cosmology we derived both the parameters in the form $H=\frac{1%
}{3}\left( H_{x}+2H_{y}\right) =\frac{m}{3t}$ and $q=-1+\frac{d}{dt}\left( 
\frac{1}{H}\right) =-1+\frac{3}{m}$. From the above two equations, we can
see that the Hubble parameter decreases with increasing cosmic time and it becomes
zero at infinite future. Also, it is clear that the deceleration parameter
takes negative values for all values of $m$ in our cosmological model. This
behavior is consistent with the current scenario of the Universe.

\section*{Acknowledgments}

We are very much grateful to the honorary referee and the editor for the
illuminating suggestions that have significantly improved our work in terms
of research quality and presentation.

\textbf{Data availability} There are no new data associated with this article.
\newline

\textbf{Declaration of competing interest} The authors declare that they
have no known competing financial interests or personal relationships that
could have appeared to influence the work reported in this paper.\newline

\end{document}